\font\twelvebf=cmbx12
\vskip 15mm
\def\Z_+={{\bf Z}_+}
\def\Z{{\bf Z}}
\def\C{{\bf C}}
\def\N{{\bf N}}
\def\R{{\bf R}}
\leftskip 50pt

{\twelvebf  Casimir invariants and characteristic
identities for $gl(\infty )$}

\vskip 1cm
\leftskip 50pt
\noindent
M.D. Gould and N.I. Stoilova\footnote{*}{Permanent address: 
Institute for Nuclear
Research and
Nuclear Energy, 1784 Sofia, Bulgaria; E-mail: stoilova@inrne.acad.bg}

\noindent
Department of Mathematics, University of Queensland, Brisbane Qld
4072,
Australia

\vskip 48pt

\noindent
A full set of (higher order) Casimir invariants for the Lie algebra 
$gl(\infty )$ is constructed and shown to be well defined in the 
category 
$O_{FS}$ generated by the highest weight (unitarizable) irreducible
representations  with only a finite number of non-zero weight 
components.
Moreover the eigenvalues of these Casimir invariants are determined
explicitly in terms of the highest weight. Characteristic identities
satisfied by certain (infinite) matrices with entries from $gl(\infty 
)$
are also determined and generalize those previously obtained for 
$gl(n)$
by Bracken and Green.$^{1,2}$

\leftskip 0pt

\vskip 32pt 

{\bf I. INTRODUCTION}

\bigskip
In recent years  infinite dimensional Lie algebras have become a 
subject
of interest in both mathematics and physics (see Refs. 3 and 4 and the
references therein).  We mention as an example, related to the topic 
of the
present article, that the Lie algebra $gl(\infty )$ and its 
completion and
central extension $a_{\infty }$ play an important role in the theory 
of
soliton equations,$^{5,6}$ string theory, two dimensional statistical
models, etc.$^7$  In addition these algebras provide an example of 
Kac-Moody
Lie algebras of infinite type.$^{3,8}$ 

In this paper, we derive  a full set of Casimir invariants for the 
infinite
dimensional general linear Lie algebra $gl(\infty )$, corresponding 
to the 
following matrix realization  (see the notation at the end
of the Introduction),

$$gl(\infty)=\{ x=(a_{ij})|\; i,j \in \N ,\; all \; but \; a \;
finite \;
number
\; of \;a_{ij} \in \C \; are \; zero \}. \eqno (1)$$
\noindent
Characteristic identities satisfied by certain infinite matrices with
entries from $gl(\infty )$ are also determined and generalize those 
obtained
by Bracken and Green$^{1,2}$ for $gl(n).$ Such identities are of 
interest and
have found applications to state labeling problems$^9$ and to the
determination of Racah-Wigner coefficients.$^{10}$ 

A basis for the Lie algebra $gl(\infty)$ is given by the Weyl 
generators $e_{ij},\; i,j\in \N,$ satisfying the commutation 
relations:

$$[e_{ij},e_{kl}]=\delta _{jk}e_{il}-\delta _{li}e_{kj}. \eqno (2)$$
The category $O $ generated by highest weight irreducible 
$gl(\infty)$ modules, 
corresponding to the "Borel" subalgebra
$$N_+=\; lin.\; env.\{e_{ij}|i<j\in \N \}, \eqno (3)$$
has been constructed in Ref. 11.
By definition  each $gl(\infty)$ module  $V \in O $ contains 
a unique (up to a multiplicative constant) vector $v_{\Lambda },$ the
highest weight vector, with the properties:
$$N_+v_{\Lambda }=0, \quad e_{ii}v_{\Lambda }=\Lambda _iv_{\Lambda },
\quad \forall i \in \N. \eqno (4)$$
The highest weight $\Lambda \equiv (\Lambda _1, \Lambda _2, \Lambda 
_3, 
\ldots )$ of $V\in O $ 
uniquely labels  the module, $V\equiv V(\Lambda )$. Moreover all
unitarizable irreducible highest weight $gl(\infty)$ modules 
$V(\Lambda ),$
corresponding to the natural conjugation operation:  
$(e_{ij})^{\dagger }=e_{ji}, \; \forall i,j \in \N ,$ have been
determined.$^{11}$ 
The module $V(\Lambda )\in O $ carries an unitarizable 
representation of $gl(\infty)$ if and only if
$$\Lambda _i-\Lambda _j \in \Z_+,
\;\; \forall i<j
\in \N, \;\; \Lambda _i\in \R , \;\; \forall i\in \N. \eqno (5)$$
In the paper we will consider the category
$O_{FS}  \subset O ,$ of modules generated by all 
unitarizable irreducible $gl(\infty )$ modules
with a finite number of non-zero highest weight components $\Lambda 
_i.$
These are modules $V(\Lambda )$ with highest weights
$$\Lambda  \equiv (\Lambda _1, \Lambda _2, \ldots , \Lambda _k ,
 0,\ldots )\equiv
(\Lambda _1, \Lambda _2, \ldots, \Lambda _k, \dot{0} ) .
 \eqno (6)$$

The paper is organized as follows. Section II gives some useful 
results on
the representations of $gl(\infty )$ with a finite number of non-zero
components of the highest weight. 
In Sec. III we construct a full set of
convergent
Casimir invariants on each module $V(\Lambda )$. 
Section IV is devoted to the computation of the eigenvalues
of these Casimir invariants for all modules from the subcategory 
$O_{FS}.$ 
In Section V  we present a derivation of the polynomial identities 
satisfied by
certain matrices with entries from $gl(\infty )$, which generalize 
those
obtained previously for $gl(n).$ 

Throughout the paper we use the following notation:

irrep(s) - irreducible representation(s);

lin.env. $\{X\}$ - the linear envelope of $X$;

$\C$ - the complex numbers;

$\R$ - the real numbers;

$\Z_+$ - all non-negative integers;

$\N$ - all positive integers;

$U(A)$ - the universal enveloping algebra of $A.$

\bigskip
{\bf II. PRELIMINARIES}
\bigskip
Denote by $H$ the Cartan subalgebra of $gl(\infty ).$ The 
space $H^*$ dual to $H$ is described by the forms 
$\varepsilon _i,\;i\in \N,$ where $\varepsilon _i:x\rightarrow 
a_{ii},$
and $x$ is given  by (1) only for diagonal $x$.  Let $ (\;,\;)$ be
the   bilinear form on $H^*$ 
 defined by $(\epsilon _i, \epsilon _j)
=\delta _{ij}.$ 
For a weight $\mu =\sum_{i=1}^{\infty }\mu _i\varepsilon _i\in H^*$ 
with
 $\mu _i$ being complex numbers we write
$\mu \equiv (\mu _1, \mu _2, \ldots , \mu _n, \ldots ).$
 The roots $\varepsilon _i-\varepsilon_j \;(i\neq j)$
 of $gl(\infty )$ are the non-zero weights of the adjoint
representation. The positive roots are given by the set:
$$\Phi ^+= \{\varepsilon _i-\varepsilon _j|
1\leq i<j\in \N \}. \eqno (7)$$
Define
$$\rho={1\over 2}\sum_{i=1}^{\infty}(1-2i)\epsilon _i. \eqno (8)$$
Let  $D_n$ be the set of $gl(\infty )$ weights: 
$$D_n=\{ \nu |\nu =(\nu_1,\ldots , \nu_n, \dot{0}), \;\; 
\nu_i \in \Z_+\} , \eqno(9)$$
and let $D_n^+\subset D_n $ be the subset of dominant weights in 
$D_n:$
$$D_n^+=\{ \nu |\nu \in D_n,\; 
(\nu , \varepsilon _i-\varepsilon _{i+1})\in \Z_+, \;\; \forall i\in 
\N\} .
 \eqno (10)$$
Denote
$$D_{FS}^+\equiv \cup _{n=1}^{\infty }D_n^+, \;\; 
D_{FS}\equiv \cup _{n=1}^{\infty }D_n. \eqno (11) $$
Note that: 

1). The irreducible $gl(\infty )$ modules $V(\Lambda )$ with highest
weights $\Lambda \in D_k^+\subset D_{FS}^+$, corresponding to the 
natural
conjugation operation,  generate 
the subcategory $O_{FS}\subset O$ of unitarizable $gl(\infty )$ 
modules 
(6);

2). Each module $V(\Lambda )$  gives rise to a unitarizable module for
the canonical subalgebra $gl(n)\subset gl(\infty )$ with generators 
$e_{ij}, \; i,j=1,\ldots ,n.$ In general $V(\Lambda )$ is a reducible
$gl(n)$ module, more precisely it is a completely reducible $gl(n)$ 
module;

3). If $\nu $ is a weight in $V(\Lambda ),$ then $\nu \in D_n,$ for 
some 
$n\in \Z_+.$

\bigskip
Let $\Lambda _n$ be the projection of the $gl(\infty )$ highest weight
$\Lambda \in D_k^+$ onto the weight space of $gl(n)$ so that, for 
$n>k,$
$$\Lambda _n=(\Lambda _1, \ldots , \Lambda _k, 0, \ldots ,0_{n})=
(\Lambda _1, \ldots , \Lambda _k, \dot{0}_{n-k}). \eqno (12)$$

{\bf Theorem 1:} {\it (i) The $gl(n)$ module 
$V_n(\Lambda ) \subset V(\Lambda ),$ $\Lambda \in D_k^+,$ cyclically 
generated by the highest 
weight vector $v_{\Lambda}^+ \in V(\Lambda ) $ is irreducible with
highest weight $\Lambda _n.$

(ii) If $v\in V(\Lambda )$ is a weight vector of weight $\nu \in D_n,$
then $v\in V_n(\Lambda ).$
} 

\bigskip
{\it Proof:} (i) The cyclic $gl(n)$ module $V_n(\Lambda )$ generated 
by 
$v_{\Lambda }^+$ is well known to be indecomposable (see for instance 
Ref. 12). The result then follows from the complete reducibility of 
$V(\Lambda )$ considered as a $gl(n)$ module.

(ii) Let $v\in V(\Lambda )$ have  weight 
$\nu\in D_n$. From the
Poincar\'e-Birkhoff-Witt theorem we may write
$$v=nv_{\Lambda }^+, \;\; n\in U(N_-), \eqno(13)$$ 
with $N_-$ the subalgebra of $gl(\infty )$ generated by  all negative
root vectors
$$N_-=lin.env.\left\{e_{ij}|i>j\in \N \right\} . \eqno(14)$$
The weight $\nu\in H^*$ has the form
$$\nu =\Lambda -\sum_{i=1}^{\infty }m_i(\varepsilon _i-\varepsilon 
_{i+1})
\eqno (15)$$
and $m_i=0$ for all but a finite number of $i.$ Since $\nu \in D_n,$  
$m_i=0$ for $i>n$ so that
$$\nu =\Lambda -\sum_{i=1}^{n }m_i(\varepsilon _i-\varepsilon _{i+1}).
\eqno (16)$$
In view of the linear independence of the simple roots $\varepsilon _i
-\varepsilon _{i+1},$ (16) implies that
$$n\in U(N_-) \cap U[gl(n)]. \eqno (17)$$
Therefore $v$ is a vector from the $gl(n)$ module $V_n(\Lambda ),$ 
$v\in V_n(\Lambda ).$ \hskip 6cm $[]$

\bigskip
Consider the $gl(\infty )$ modules $V(\Lambda )$ and $V(\mu ),$
with highest weights $\Lambda \in D_k^+$ and $\mu \in D_l^+,$ 
respectively.
Take the tensor product of them 
$$V(\Lambda )\otimes V(\mu ), \eqno (18)$$ 
and suppose that $v_{\nu }^+$ is a $gl(\infty )$ highest weight 
vector in
(18). Then for some $n, \quad \nu \in D_n^+$ so that $v_{\nu }^+$ is a
linear combination of vectors of the form
$$v\otimes w, 
\eqno (19)$$
where $v$ and $ w$ have weights in $D_n$. 
{\it Theorem 1} then implies that $v\in V_n(\Lambda ), \;\; w\in 
V_n(\mu ).$
Therefore 
$$v_{\nu }^+ \in V_n(\Lambda )\otimes V_n(\mu ). \eqno (20)$$
Since  $\Lambda $ has $k$ and $\mu $ has $l$ non-zero components,  
then $\nu $ can have at
most $k+l$ non-zero components, so that $n\leq k+l.$ Hence w.l.o.g. 
we may
take $n=k+l.$
Thus if $v_{\nu }^+$ is a $gl(\infty )$ highest weight vector in  (18)
then
$$v_{\nu }^+\in V_n(\Lambda )\otimes V_n(\mu ), \quad n=k+l, \eqno 
(21)$$
is a $gl(n)$ highest weight vector. Conversely, given a $gl(n)$ 
highest
weight vector
$$v_{\nu }^+\in V_n(\Lambda )\otimes V_n(\mu ), \quad n=k+l, $$
we have
$$e_{ij}v_{\nu }^+=0,\quad \forall i<j=1,\ldots , n, $$
while
$$e_{ij}v_{\nu}^+=0, \quad \forall j>n,$$
since all weights in $V(\Lambda )$ and $V(\mu )$ have entries in 
$\Z_+.$ Therefore $v_{\nu }^+$ must be a $gl(\infty )$ highest weight
vector. $V_n(\Lambda )\;$ and $V_n(\mu )$ are $gl(n)$ irreducible 
modules
with highest weights $\Lambda _n$ and $\mu _n$ respectively. 
For their tensor product decomposition we write
$$V_n(\Lambda )\otimes V_n(\mu )\equiv V(\Lambda _n)\otimes V(\mu _n)=
\oplus _{\nu }m_{\nu }V(\nu _n)\equiv 
\oplus _{\nu }m_{\nu }V_n(\nu ), \eqno (22)
$$
where $\nu \equiv (\nu_n, \dot{0}).$

Hence we have proved:

{\bf Theorem 2:} {\it The irreducible $gl(n)$ module decomposition}
$$V_n(\Lambda )\otimes V_n(\mu )=\oplus_{\nu } m_{\nu }V_n(\nu ), 
\eqno (23)$$
{\it implies the $gl(\infty )$ irreducible module decomposition}
$$V(\Lambda )\otimes V(\mu )=\oplus_{\nu } m_{\nu }V(\nu ),
 \eqno (24)$$
$where \;\; \Lambda \in D_k^+,\; \mu \in D_l^+,\;\;  n=k+l.$

\hskip 15.5cm $[]$

\bigskip
{\bf III. CONSTRUCTION OF CASIMIR INVARIANTS }

\bigskip
An obvious invariant for $gl(\infty )$ is the first order invariant

$$I_1=\sum_{i=1}^{\infty }e_{ii}. \eqno(25)$$
However, it is not  clear how to 
construct appropriate higher order invariants for 
$gl(\infty )$ . Let us therefore consider
the second order invariant $I_2^{(n)}$ of $gl(n):$

$$
\eqalignno{
& I_2^{(n)}=\sum_{i,j=1}^ne_{ij}e_{ji}=\sum_{i=1}^n\sum_{j<i=1}^n
e_{ij}e_{ji}+\sum_{i=1}^n\sum_{j>i=1}^n
e_{ij}e_{ji}+\sum_{i=1}^n e_{ii}^2 =2\sum_{i=1}^n\sum_{j<i=1}^n
e_{ij}e_{ji}+\sum_{i=1}^n\sum_{j>i=1}^n
(e_{ii}-e_{jj}) & \cr  
&&\cr
& +\sum_{i=1}^n e_{ii}^2 =
2\sum_{i=1}^n\sum_{j<i=1}^n
e_{ij}e_{ji}+\sum_{i=1}^n 
(n+1-2i)e_{ii}+\sum_{i=1}^n e_{ii}^2 =2\sum_{i=1}^n\sum_{j<i=1}^n
e_{ij}e_{ji}+\sum_{i=1}^n e_{ii}(e_{ii}+1-2i) +nI_1^{(n)},
& \cr
&& (26) \cr
}
$$
where $I_1^{(n)}\equiv \sum_{i=1}^ne_{ii}$ is the first order 
invariant of
$gl(n).$ Due to the last term in (26) the $gl(n)$ second order 
invariant
diverges as $n\rightarrow \infty .$ Eliminating the last term in (26) 
(the rest 
of the expression is also an invariant) and taking the limit 
$n\rightarrow \infty $ one obtains the following quadratic Casimir 
for 
$gl(\infty )$:
$$I_2=2\sum_{i=1}^{\infty }\sum_{j<i}^{\infty }
e_{ij}e_{ji}+\sum_{i=1}^{\infty } e_{ii}(e_{ii}+1-2i), \eqno(27) $$
which is convergent (see formula (36)) on the category $O_{FS}$ 
of irreps considered. On 
$V(\Lambda ), \;\Lambda \in D_k^+,\;\; I_2$ takes constant value
$$\chi _{\Lambda } (I_2)=\sum_{i=1}^k \Lambda _i (\Lambda _i+1-2i)
=(\Lambda, \Lambda+2\rho). \eqno (28)$$
This construction suggests  how to proceed to the higher order
invariants of $gl(\infty ).$

To begin with we introduce the characteristic matrix
$$A_i^{\;j}=e_{ji}. \eqno (29)$$
This matrix, in fact, arises naturally in the context of 
characteristic
identities, to be discussed in Sec. V. Powers of the matrix $A$ are 
defined
recursively by
$$\left( A^m \right) _i^{\;j}=\sum_{k=1}^{\infty }A_i^{\;k}
(A^{m-1})_k^{\;j}, \quad\quad 
[(A^0)_i^{\;j}\equiv \delta _{ij}]. \eqno (30)$$

 Using  induction and the $gl(\infty )$
commutation relations (2) one obtains:

{\bf Proposition 1:} 

$$[e_{kl}, (A^m)_i^{\;j}]=\delta _{jl}(A^m)_i^{\;k}-\delta _{ik}
(A^m)_l^{\;j}. \eqno (31)$$
\hskip 16 cm $[]$

\noindent
Therefore the matrix traces
$$tr(A^m)\equiv \sum_{i=1}^{\infty}(A^m)_i^{\;i} \eqno (32)$$
are formally Casimir invariants. They are, however, divergent except 
for 
$m=1$ in
which case we obtain the first order invariant (25). The purpose of 
the
present investigation  is to construct a full set of Casimir 
invariants which
are well defined and convergent on the category $O_{FS}.$

The following is the main result of the paper:

{\bf Theorem 3:} {\it The Casimir invariants defined recursively by}
$$
\eqalignno{
& I_1=\sum_{i=1}^{\infty} A_i^{\;i}=tr(A); &  \cr
& I_m=\sum_{i=1}^{\infty} \left [(A^m)_i^{\;i}-I_{m-1}\right]=
tr\left[ A^m-I_{m-1}\right] & (33) \cr
}
$$
{\it form a full set of convergent Casimir invariants on each module }
$V(\Lambda )\in O_{FS}$.\hskip 4cm $[]$

Observe first that the $I_m$ so defined (33) are indeed Casimir 
invariants
(see {\it Proposition 1}). It remains to prove they are convergent on 
the category
$O_{FS}.$ We will do this by induction. It is constructive to 
consider 
first the case $m=2:$ 
$$
\eqalignno{
& I_2\equiv \sum_{j=1}^{\infty }\left[(A^2)_j^{\;j}-I_1\right]=
\sum_{j=1}^{\infty }\left[ \sum_{i=1}^{\infty }e_{ij}e_{ji}-
I_1\right]= 
 \sum_{j=1}^{\infty }\left[ \sum_{i>j}^{\infty }
e_{ij}e_{ji}+\sum_{i<j}^{\infty }
e_{ij}e_{ji}+e_{jj}^2-I_1\right] & \cr
& \;\; = \sum_{j=1}^{\infty }\left[ 2\sum_{i>j}^{\infty }
e_{ij}e_{ji}+\sum_{i<j}^{\infty }
(e_{ii}-e_{jj})+e_{jj}^2-I_1\right]=
\sum_{j=1}^{\infty }\left[ 2\sum_{i>j}^{\infty }
e_{ij}e_{ji}+e_{jj}(e_{jj}-j+1)+
\sum_{i<j}^{\infty }
e_{ii}-I_1\right] & \cr
& \;\; =\sum_{j=1}^{\infty }\left[ 2\sum_{i>j}^{\infty }
e_{ij}e_{ji}+e_{jj}(e_{jj}-j)-
\sum_{i>j}^{\infty }
e_{ii}\right]=2\sum_{j=1}^{\infty }\sum_{i>j}^{\infty }
e_{ij}e_{ji}+
\sum_{j}^{\infty }
e_{jj}(e_{jj}-2i+1), & \cr
&& (34) \cr
}
$$
which agrees with the definition (27).

Now let ${ \it v}\in V(\Lambda ),\;\Lambda \in D_k^+,\;$ 
 be an arbitrary
weight  vector. Then the weight of $ {\it v}$ has the form
$$\nu =(\nu _1,\nu _2, \ldots , \nu _r, \dot{0}), \eqno (35)$$
so that $\sum_{i=1}^r\nu _i=\sum_{i=1}^k\Lambda _i=\chi _{\Lambda 
}(I_1).$
Note that
$$A_i^{\;j}{\it v}=e_{ji}{\it v}=0, \;\; \; \forall i>r, \eqno (36)$$
and that the second order invariant $I_2$ is convergent on each
$V(\Lambda )\in O_{FS} $ [c.f. formula (27)].

Applying {\it Proposition 1} and (36), for $i>r$ one  obtains
$$
\eqalignno{
& (A^m)_i^{\;i}{\it v}=\sum_{j=1}^{\infty }A_i^{\;j}(A^{m-
1})_j^{\;i}{\it v}
=\sum_{j=1}^{\infty }e_{ji}(A^{m-1})_j^{\;i}{\it v} & \cr
& =
\sum_{j=1}^{\infty }\left\{ \left[ (A^{m-1})_j^{\;j}-(A^{m-
1})_i^{\;i}\right]
 {\it v}+
(A^{m-1})_j^{\;i}e_{ji}{\it v} \right\}
 =\sum_{j=1}^{\infty }\left[ (A^{m-1})_j^{\;j}-(A^{m-
1})_i^{\;i}\right] {\it
v}.
& (37) \cr
}
$$
In particular for the case $m=2$ we have

$$(A^2)_i^{\;i}{\it v}=\sum_{j=1}^{\infty }\left[ 
A_j^{\;j}-A_i^{\;i}\right] {\it v}=\sum_{j=1}^{\infty }e_{jj}{\it v}
=I_1{\it v}, \;\; 
\forall i>r \eqno (38)$$
so that
$$\left( (A^2)_i^{\;i}-I_1\right) {\it v}=0,\; \forall i>r, \eqno 
(39)$$
which is another proof for the convergence of $I_2.$
More generally

{\bf Proposition 2:} {\it For any weight vector} ${ \it v}\in 
V(\Lambda ),$ 
{\it and $m\in \N$
there exist 
$r\in \N$ such that}
$$\left((A^m)_i^{\;i}-I_{m-1}\right){\it v}=0,\;\;\forall i>r. 
\eqno(40)$$

{\it Proof:} We proceed by induction and assume $v$ has weight $\nu $ 
as in
(35). Formula (40) is valid for $m=2$ (39). 
Assuming the result is
true for a given $m$, i.e.
$$(A^m)_i^{\;i} {\it v}=I_{m-1}{\it v},\;\;\forall i>r$$
we have (see (37))
$$(A^{m+1})_i^{\;i}{\it v}=
\sum_{j=1}^{\infty }\left[ (A^{m})_j^{\;j}-(A^{m})_i^{\;i}\right] 
{\it v}=
\sum_{j=1}^{\infty }\left[ (A^{m})_j^{\;j}-I_{m-1}\right] {\it 
v}=I_m{\it v},
\quad \forall i>r, \eqno (41)$$
which proves (40). 
\hskip 13 cm $[]$

\bigskip

$I_m$ (33) is convergent on each $V(\Lambda )$ for $m=2.$ Assume it is
convergent and well defined on $ V(\Lambda )$ 
for a given
$m.$ Then, with $v$ as in (40), we have
$$I_{m+1}{\it v}\equiv \sum_{i=1}^{\infty}\left[ (A^{m+1})_i^{\;i}-I_m
\right] {\it v}=\sum_{i=1}^{r}\left[ (A^{m+1})_i^{\;i}-I_m
\right] {\it v}=\sum_{i=1}^{r}(A^{m+1})_i^{\;i}{\it v} -rI_m
 {\it v}, \eqno (42)$$
so that $I_{m+1}$ is convergent and well defined on $V(\Lambda ).$

This completes the (inductive) proof of {\it Theorem 3}. 

\bigskip
In the next Section we will obtain an explicit eigenvalue formula for 
these
invariants.

\bigskip
{\bf IV. EIGENVALUE FORMULA FOR CASIMIR INVARIANTS}

\bigskip
In this section we apply our previous results to evaluate the 
spectrum of
the invariants (33).

Let $v\in V(\Lambda ),$ be an arbitrary vector of weight 
$\nu =(\nu_1, \ldots , \nu _r, \dot{0}).$ Then, keeping in mind {\it
Proposition 1,} the fact that $(A^{m-1})_k^{\;j}$ has weight 
$\varepsilon
_j-\varepsilon _k$ under the adjoint representation of $gl(\infty )$ 
and
that all vectors of $V(\Lambda )$ have weight components in $\Z_+,$ 
we must
have for $j\leq r$
$$(A^{m-1})_k^{\;j}v=0, \quad \forall k>r. \eqno (43)$$
Therefore
$$(A^m)_i^{\;j}v=\sum_{k=1}^{\infty }A_i^{\;k}(A^{m-1})_k^{\;j}v
=\sum_{k=1}^{r }A_i^{\;k}(A^{m-1})_k^{\;j}v. \eqno (44)$$
Proceeding recursively  we may therefore write
$$(A^m)_i^{\;j}v=(\bar{A}^{m})_i^{\;j}v, \quad \forall i,j=1,
\ldots, r, \eqno (45)$$
where $(\bar{A})_i^{\;j}=e_{ji}, \;\; \forall i,j=1,\ldots , r,$ is 
the
$gl(r)$ characteristic matrix, and the powers of the matrix $\bar{A}$ 
are
defined by (30) with $i,j,k=1,\ldots ,r$ and $\bar{A}$ instead of $A.$
It follows then that the formula (42) can be written as:
$$I_{m}{\it v}= \sum_{i=1}^{r}\left[ (\bar{A}^{m})_i^{\;i}-I_{m-1}
\right] {\it v}=\left[ I_m^{\;(r)}-rI_{m-1}
\right] {\it v}, \eqno (46)$$
with 
$$I_m^{\;(r)}=\sum_{i=1}^r(\bar{A}^m)_i^{\;i}, \eqno(47)$$
 being the 
$m^{th}$ order invariant of $gl(r).$
Formula (46) is valid $\forall m\in \N,$ which gives a recursion 
relation
for the $I_m$ with initial condition
$$I_1v=\chi _{\Lambda }(I_1)v. \eqno (48)$$
In particular it follows from (46) that the invariants $I_m$ are 
certainly
convergent on all weight vectors $v\in V(\Lambda ).$

To determine the eigenvalues of $I_m$ let $v=v_{\Lambda }^+$ be the 
highest
weight vector of the unitarizable module $V(\Lambda )$ 
and let
$$\Lambda =(\bar{\Lambda }, \dot{0})\in D_k^+, \quad \bar{\Lambda 
}\equiv 
(\Lambda _1, \ldots , \Lambda _k). \eqno (49)$$
Then for the eigenvalues of the $I_m$ one obtains the recursion 
relation
(see (46)): 
$$\chi_{\Lambda }(I_m)=\chi_{\bar{\Lambda }}(I_m^{\;(k)})-
k\chi _{\Lambda }(I_{m-1}), \quad \chi_{\Lambda }(I_1)=\sum_{i=1}^k
\Lambda _i, \eqno (50)$$
where $\chi_{\bar{\Lambda }}(I_m^{\;(k)})$ is the eigenvalue of the 
$m^{th}$
order invariant (47) of $gl(k)$ on the irreducible $gl(k)$ module with
highest weight $\bar{\Lambda };$ the latter is given explicitly by
$$\chi_{\bar{\Lambda }}(I_m^{\;(k)})=\sum_{i=1}^k
\alpha _i^m \prod_{j\neq i=1}^k\left(
{\alpha _i-\alpha _j+1\over{\alpha _i-\alpha _j}}\right), \eqno (51)$$
where
$$\alpha _i=\Lambda _i+1-i. $$
We thereby obtain for the eigenvalues of the Casimir invariants $I_m$
$$\chi _{\Lambda }(I_m)=\sum_{i=1}^kP_m(\alpha _i)
\prod_{j\neq i=1}^k\left({\alpha _i-\alpha _j+1\over{\alpha _i-\alpha 
_j}}
\right),
 \eqno (52)$$
for suitable polynomials $P_m(x)$ which, from Eq. (50), satisfy the 
recursion
relation
$$P_m(x)=x^m-kP_{m-1}(x), \quad P_1(x)=x. \eqno (53)$$
In particular
$$
\eqalignno{
& P_2(x)=x^2-kx=x{x^2-k^2\over{x+k}}; & (54a) \cr
& P_3(x)=x^3-k(x^2-kx)=x{x^3+k^3\over{x+k}}, & (54b) \cr
}
$$
and more generally, it is easily established by induction that
$$P_m(x)=x{x^m-(-1)^mk^m\over{x+k}}. \eqno (55)$$ 
Thus we have

{\bf Theorem 4:} {\it The eigenvalues of the Casimir invariants $I_m$ 
(33),
 on the
irreducible unitarizable $gl(\infty )$ module $V(\Lambda )$, 
$\Lambda \in D_k^+$
are given by }
$$\chi_{\Lambda }(I_m)=\sum_{i=1}^k\alpha _i\left({\alpha _i^m+
(-1)^{m+1}k^m\over{\alpha _i+k}}\right) 
\prod_{j\neq i}^k\left( {\alpha _i-\alpha _j+1\over{\alpha _i-\alpha 
_j}}
\right), \quad
where \;\;
\alpha _i=\Lambda _i+1-i. \eqno (56)$$
\hskip 16cm $[]$

\smallskip
{\bf V. POLYNOMIAL IDENTITIES}

\bigskip
Let $\Delta $ be the comultiplication on the enveloping algebra
$U[gl(\infty )]$ of $gl(\infty )$ ($\Delta (e_{ij})=e_{ij}\otimes 1+
1\otimes e_{ij}, \;i,j\in \N ,$ with $1$ being the unit in  
$U[gl(\infty )]$).
Applying $\Delta $ to the second order Casimir invariant (27) of
$gl(\infty )$ we obtain:
$$\Delta (I_2)=I_2\otimes 1+1\otimes I_2+2\sum_{i,j=1}^{\infty }
e_{ij}\otimes e_{ji}. \eqno (57)$$
Therefore
$$\sum_{i,j=1}^{\infty }e_{ij}\otimes e_{ji}=
{1\over 2}\left[\Delta (I_2)-I_2\otimes 1-1\otimes I_2\right]. \eqno 
(58) $$
Denote by $\pi_{\varepsilon _1}$ the irrep of $gl(\infty )$ afforded 
by 
$V(\varepsilon _1).$ The weight spectrum for the vector module 
$V(\varepsilon _1)$ consists of all weights $\varepsilon _i, \;
i=1,2, \ldots , $ each occurring exactly once. Denote by $E_{ij}, \; 
i,j\in
\N$ the generators on this space
$$\pi_{\varepsilon _1}(e_{ij})=E_{ij}, \eqno (59)$$
with $E_{ij}$ an elementary matrix.

As for the algebra $gl(n)$, we introduce the characteristic matrix
$$A=\sum_{i,j=1}^{\infty }\pi_{\varepsilon _1}(e_{ij})e_{ji}=
\sum_{i,j=1}^{\infty }E_{ij}e_{ji}=
{1\over 2}(\pi_{\varepsilon _1}\otimes 1)\left[\Delta (I_2)-
I_2\otimes 1-1\otimes I_2\right]. \eqno (60)$$ 
Therefore $A$ is the infinite matrix introduced in Sec. III (see (29))
and the entries  of the matrix powers $A^m$ are given recursively by 
(30). 
We will show that the characteristic matrix satisfies a polynomial 
identity
acting on the $gl(\infty )$ module $V(\Lambda ), \; \Lambda \in 
D_k^+.$ 
Let $\pi_{\Lambda }$ be the representation afforded by $V(\Lambda ).$
From Eq.
(60), acting on $V(\Lambda )$ we may interpret $A$ as an invariant
operator on the tensor product module 
$V(\varepsilon _1)\otimes V(\Lambda ):$
$$A\equiv {1\over 2}(\pi_{\varepsilon _1}\otimes \pi_{\Lambda })
\left[ \Delta (I_2)-I_2\otimes 1-1\otimes I_2\right]. \eqno (61)$$
From {\it Theorem 2,} we have for the tensor product decomposition
$$V(\varepsilon _1)\otimes V(\Lambda )=
\oplus_{i=1}^{k+1} {^ \prime}  \;
V(\Lambda +\varepsilon _i),\eqno(62)$$ 
where the prime signifies that  it is necessary to retain only those
summands for which $\Lambda +\varepsilon _i\in D_{FS}^+.$ Therefore 
on each
$gl(\infty )$ module $V(\Lambda +\varepsilon _i)$ in (62), $A$ takes
the eigenvalue
$${1\over 2}\left[ \chi_{\Lambda +\varepsilon_i}(I_2)-
\chi_{\varepsilon_1}(I_2)-\chi_{\Lambda }(I_2)\right]=
{1\over 2}\left[(\Lambda +\varepsilon _i, \Lambda +\varepsilon 
_i+2\rho )-
(\varepsilon _1, \varepsilon _1+2\rho )-(\Lambda , \Lambda +2\rho 
)\right]
=\Lambda _i+1-i \eqno (63)$$  
(see {\it Theorem 4).} Thus we have

{\bf Theorem 5:} {\it On each $gl(\infty )$ module $V(\Lambda )$, 
$\Lambda \in D_k^+$ the
characteristic matrix satisfies the polynomial identity}
$$\prod_{i=1}^{k+1}\left( A-\alpha _i\right)=0, 
 \eqno(64)$$
{\it with $\;\alpha _i=\Lambda _i+1-i\;$ the characteristic roots.}
\hskip 9cm $[]$

\noindent
The characteristic identities (64) are the $gl(\infty )$ counterpart 
of
the polynomial identities encountered for $gl(n)$ by Bracken and Green
$^{1,2}$ (more precisely their adjoint identities).
It is worth noting, in view of the decomposition (62), that these 
identities may
frequently be reduced. Some reduced identities  are indicated below 
for
certain choices $\Lambda \in D_{FS}^+$ of the $gl(\infty )$ highest
weight:
$$
\eqalignno{
& \Lambda =(\dot{1}_k, \dot{0}): \quad\;\;\; (A-1)(A+k)=0; & (65a) \cr
& \Lambda =(k, \dot{0}): \quad\quad\; (A+1)(A-k)=0; & (65b) \cr
& \Lambda =(\dot{p}_k, \dot{q}_l, \dot{0}): \;\; (A-p)(A+k-q)
(A+k+l)=0, \;\; p<q. & (65c)\cr
}
$$
{\it Note:} Sometimes the characteristic and reduced identities are 
the
same; for instance in (65b) the reduced identity coincides with the
characteristic identity. This is in stark contrast to the 
characteristic
identities for $gl(n).$

\bigskip
More generally, having in mind (58), introduce a characteristic 
matrix 
$$A_{\Lambda }=\sum_{i,j=1}^{\infty }\pi _{\Lambda }(e_{ij})e_{ji}=
{1\over 2}(\pi _{\Lambda }\otimes 1)\left[ \Delta (I_2)-I_2\otimes 1-
1\otimes I_2\right], \eqno(66)$$  
corresponding to any irrep $\pi _{\Lambda }$ of $gl(\infty )$ 
afforded by 
$V(\Lambda ), \; \Lambda \in D_k^+.$ In a suitably chosen basis for 
$V(\Lambda )\;$ $A_{\Lambda }$ is an infinite matrix with entries
$$\left(A_{\Lambda }\right)_{\alpha }^{\; \beta}=
\sum_{i,j=1}^{\infty }\pi _{\Lambda }(e_{ij})_{\alpha  \beta}e_{ji}. 
\eqno(67)$$
Acting on an irreducible $gl(\infty )$ module 
$V(\mu ), \; \mu \in D_l^+,\;$ 
$A_{\Lambda }$ may be regarded as an invariant operator on the tensor
product module $V(\Lambda ) \otimes V(\mu ):$
$$A_{\Lambda }\equiv {1\over 2}(\pi _{\Lambda }\otimes \pi _{\mu})
\left[ \Delta (I_2)-I_2\otimes 1-
1\otimes I_2\right]. \eqno(68)$$
Now applying {\it Theorem 2,}  the decomposition of the tensor product
space $V(\Lambda )\otimes V(\mu )$ is given by the $gl(k+l)$
branching rule
$$V_n(\Lambda )\otimes V_n(\mu )=\oplus_{\nu } m_{\nu }V_n(\nu ), 
\eqno(69)$$ 
with $n=k+l.$ Let $\{\lambda _n^i \}_{i=1}^d$ be the set of distinct 
weights
in the $gl(n)$ module $V_n(\Lambda ).$ Then the allowed highest 
weights 
$\nu _n$ occurring in the decomposition (69) are of the form 
$\nu _n=\mu _n+\lambda _n^i,$ for some $i.$ It follows that on 
$V(\nu ),\;\; \nu =(\nu _n, \dot{0}),$ the matrix $A_{\Lambda }$ takes
the constant values
$$\alpha _{\Lambda ,i}={1\over 2}\left[ \chi _{\mu +\lambda _i }
(I_2)-\chi _{\Lambda }(I_2)-\chi _{\mu }(I_2)\right] =
{1\over 2}\left[ (\lambda _i,\lambda _i+2(\mu +\rho))-(\Lambda , 
\Lambda +
2\rho )\right], \quad \lambda _i=(\lambda _n^i, \dot{0}), \eqno (70) 
$$
which are the characteristic roots of the matrix $A_{\Lambda }.$ Thus 
we
have

{\bf Theorem 6:} {\it On the irreducible $gl(\infty )$ module $V(\mu 
),\;
\mu \in D_{FS}^+,$ the characteristic matrix $A_{\Lambda }$ satisfies 
the
polynomial identity}
$$\prod_{i=1}^d(A_{\Lambda }-\alpha _{\Lambda ,i})=0. \eqno(71)$$ 
\hskip 16cm $[]$

\bigskip
\noindent
These identities are obvious generalizations of those of {\it Theorem 
5}
(see (64)). Note, in this case, that Eq. (69) implies the reduced 
identity
satisfied by the matrix $A_{\Lambda }$ on the $gl(\infty )$ module
$V(\mu )$  given by
$$\prod_{\nu}(A_{\Lambda }-\alpha _{\nu })=0, \eqno (72) $$
where now
$$\alpha _{\nu }={1\over 2}\left[ (\nu , \nu +2\rho )-(\Lambda , 
\Lambda
+2\rho )-(\mu , \mu +2\rho )\right]. \eqno (73)$$

\bigskip
Casimir invariants for the infinite dimensional general linear Lie 
algebra
have been obtained explicitly, and their eigenvalues on  any 
irreducible
highest weight unitarizable representation with a finite number of 
non-zero
weight components computed. With the help of the second order Casimir
invariant we have obtained characteristic identities for the Lie 
algebra
$gl(\infty )$ which are a generalization of those for $gl(n).$

Is is well known that the invariants of finite dimensional Lie 
algebras play
an important role in their representation theory. However, for the 
infinite
dimensional Lie algebras  corresponding full sets of Casimir 
invariants have not  yet been determined. The present paper is a step 
in solving 
this problem.

\bigskip
{\bf ACKNOWLEDGMENTS}

\smallskip
One of us (N.I.S.) is grateful  for the kind
invitation to work in the mathematical physics group at the 
Department of Mathematics
in University of Queensland.  The
work was supported by the Australian Research Council and by the
Grant $\Phi -416$ of the Bulgarian Foundation for Scientific
Research.

\bigskip 
\noindent
{\bf References}
 
\vskip 12pt
{\settabs \+  $^{11}$& I. Patera, T. D. Palev, Theoretical
   interpretation of the experiments on the elastic \cr
   %sample line,  see p. 232 of the Texbook.

\+ $^1$&  A.J. Bracken and H.S. Green, J. Math. Phys. {\bf 12}, 2099 
(1971).
\cr

\+ $^2$  & H.S. Green, J. Math. Phys. {\bf 12}, 2106 (1971). \cr
 
\+ $^3$  & V.G. Kac, {\it Infinite dimensional Lie algebras} {\bf
44} (Cambridge University Press,
Cambridge, 1985). \cr
 
\+ $^4$ & V.G. Kac  and A.K. Raina, {\it Bombay lectures on highest 
weight
         representations of infinite}  \cr
\+     &  {\it dimensional Lie algebras} in Advanced Series in
          Mathematics {\bf 2} (World Scientific, Singapore, 1987). \cr
 
\+ $^5$ & E. Date, M. Jimbo, M. Kashiwara and T. Miwa, {\it 
Transformation
         groups for soliton equations} \cr
\+     & Publ RIMS Kyoto Univ {\bf 18}, 1077 (1982). \cr

\+ $^6$ & M. Sato, {\it Soliton equations as dynamical systems on 
infinite
dimensional Grassmann manifolds}, \cr
\+    &  RIMS Kokyoroku, {\bf 439}, 30 (1981). \cr

\+ $^7$ & P. Goddard and D. Olive, Int. J. Mod. Phys. {\bf A 1}, 303
         (1986). \cr

\+ $^8$ & B. Feigin and D. Fuchs,  {\it
 Representations of the Virasoro
algebra} in 
         Representations of infinite \cr
\+     & dimensional Lie groups and Lie algebras (New York: Gorgon and
         Breach, 1989). \cr

\+ $^9$ & S.A. Edwards and M.D. Gould, J. Phys. {\bf A19}, 1523; 
1531; 1537 
(1986). \cr

\+ $^{10}$ & M.D. Gould, J.Math.Phys. {\bf 21}, 444 (1980); {\bf 22}, 
15;
2376 (1981); {\bf 23}, 1944 (1986). \cr

\+ $^{11}$ & T.D. Palev, J. Math. Phys. {\bf 31}, 579 (1990). \cr

\+ $^{12}$ & J.E. Hamphreys, {\it Introduction to Lie Algebras and
Representation Theory} (Springer, Berlin, 1972). \cr

\end